%% file: aber-interp-v2arxiv.tex
\documentclass[11pt,a4paper]{article}
\pdfoutput=1
\usepackage{jcappub}
\usepackage{ifthen}
\usepackage[T1]{fontenc}
\usepackage[utf8]{inputenc}
\usepackage[english]{babel}
\usepackage{ifthen}
\usepackage[usenames,dvipsnames,svgnames]{xcolor}
\usepackage{amsmath,amssymb}

\usepackage{cancel}
\usepackage[colorinlistoftodos]{todonotes}
\usepackage{import}

\usepackage{my-own-commands}
\usepackage{natbib}

\newcommand{\dd}{\textrm{d}}

\title{Interpreting the CMB aberration and Doppler measurements: boost or intrinsic dipole?}

\def\barcel{Departament de F\'isica Fondamental i Institut de Ci\'encies del Cosmos, Universitat de Barcelona, Mart\'i i Franqu\'es 1, E-08028 Barcelona, Spain}
\def\riodej{Instituto de F\'\i sica, Universidade Federal do Rio de Janeiro, 21941-972, Rio de Janeiro, RJ, Brazil}

\author[a]{Omar Roldan,}
\author[b]{Alessio Notari}
\author[a]{and Miguel Quartin}

\affiliation[a]{\riodej}
\affiliation[b]{\barcel}

\abstract{
The aberration and Doppler coupling effects of the Cosmic Microwave Background (CMB) were recently measured by the Planck satellite. The most straightforward interpretation leads to a direct detection of our peculiar velocity $\beta$, consistent with the measurement of the well-known dipole. In this paper we discuss the assumptions behind such interpretation. We show that Doppler-like couplings appear from two effects: our peculiar velocity and a second order large-scale effect due to the dipolar part of the gravitational potential.  We find that the two effects are exactly degenerate but \emph{only} if we assume second-order initial conditions from single-field Inflation. Thus, detecting a discrepancy in the value of $\beta$ from the dipole and the Doppler couplings implies the presence of a primordial non-Gaussianity. We also show that aberration-like signals likewise arise from  two independent effects: our peculiar velocity and lensing due to a first order large-scale dipolar gravitational potential, independently on Gaussianity of the initial conditions.  In general such effects are not degenerate and so  a discrepancy between the measured $\beta$ from the dipole and aberration could be accounted for by a  dipolar gravitational potential. Only through a fine-tuning of the radial profile of the potential it is possible to have a complete degeneracy with a boost effect. Finally we discuss that we also expect other signatures due to integrated second order terms, which may be further used to disentangle this scenario from a simple boost.}

\keywords{CMB theory, CMB aberration, CMB dipole, CMB second-order perturbations}

\begin{document}

\maketitle

\section{Introduction}

Although we have attained a high-level of both precision and accuracy in our measurements of the Cosmic Microwave Background (CMB), an observation of the intrinsic cosmological dipole still eludes us. The reason is because its most straightforward effect -- a dipole anisotropy in temperature -- is degenerate with the standard Doppler effect due to our peculiar motion. For typical cosmic velocities (hundreds of km/s) the Doppler effect is much higher than the simplest expectation for the magnitude of the intrinsic dipole, to wit a $\sim 10^{-5}$ temperature anisotropy, which is what we observe on the other multipoles. Assuming that the intrinsic component is negligible the total measured dipole can be converted into a measurement of our peculiar velocity. To this end, use is made of the observed monopole $T_0=(2.7255 \pm 0.0006)$K~\cite{Fixsen:2009ug} and the observed dipole $T_{\rm dip}= (3364.5 \pm 2.0) \mu$K~\cite{Adam:2015vua} to derive our peculiar velocity $\beta \equiv | \vo|/c = (1.2345 \pm 0.0007) \, 10^{-3}$, a value that has changed little in the past 20 years~\cite{Lineweaver:1996xa}.

Such high precision is not necessarily met by a similar accuracy because in principle the dipole might have other contributions, as discussed for instance in~\cite{Grishchuk:1978,Turner:1991dn,Langlois:1995ca,Langlois:1996ms}.
This uncertainty has led to a search of other observables that could provide independent measurements of our peculiar velocity. The most straightforward one is to look for a dipole in other distant sources, such as the cosmic infrared background (see, e.g.~\cite{Fixsen:2011qk}). Even though this avenue is yet to produce precise results for diffuse sources, for galaxy surveys it may yield a $5\sigma$ detection in the next decade~\cite{Yoon:2015lta}. Another possibility is to look for secondary effects on the CMB itself that a peculiar velocity might induce. The simplest of these is an effect of order the $\ell$-th power of the dipole on a multipole $\ell$ (i.e., a $10^{-6}$ Doppler quadrupole and $10^{-9}$ Doppler octupole). However, these quantities are smaller than the primordial fluctuations, and cannot be used to measure $\beta$. The temperature dipole also shows up as a frequency-dependent effect at second-order, appearing in the so-called $y$-channel of the CMB. But this effect has been shown not to distinguish between an intrinsic and Doppler dipole~\cite{Kamionkowski:2002nd,Notari:2015daa}.

Another secondary effect is aberration, as originally discussed in~\cite{Challinor:2002zh}: in a boosted frame, the CMB anisotropies are displaced by an amount which is (to first order in $\beta$) dipolar modulated. The actual effect on the CMB spectrum is to introduce  $10^{-8}$ couplings among neighboring multipoles at all scales  both due to Doppler and aberration. It was later realized in~\cite{Kosowsky:2010jm,Amendola:2010ty} that this could be observed by the Planck satellite, and at over the $5\sigma$ confidence level~\cite{Notari:2011sb}. Such a measurement was later performed by Planck, although systematic errors in practice limited the significance to around $3\sigma$: $\beta = (1.28 \pm 0.46) \, 10^{-3}$, and with a direction consistent with the dipole~\cite{Aghanim:2013suk}.

This measurement of the Doppler and aberration couplings seem at first to confirm the hypothesis that the dipole is due to a Lorentz boost of the CMB. However, so far a more in-depth discussion of the interpretation of this measurement has been lacking. We aim to address this gap in the present paper. In particular, it is not obvious a priori if one could not also mimic the same effects with a large scale dipolar gravitational potential. Such a potential could clearly reproduce the dipole at linear order, but could it also induce similar couplings between multipoles at second order? The scope of the present paper is to check whether a dipolar potential can produce both Doppler-like and aberration-like couplings exactly in the same way as a peculiar velocity does.

This paper is organized as follows. We start by revising second order perturbation theory in section~\ref{sec:2nd-order}, adapting the formalism to make the aberration effect more explicit. In section \ref{sec:abe+len} we show that although an aberration-like coupling could in principle be produced by lensing due to a gravitational potential without a peculiar velocity, a very special type of potential would be required. If one dismisses such a possibility, the aberration measurements are indeed a direct probe of our peculiar velocity. In section~\ref{sec:doppler} we show instead that the Doppler-like couplings are more naturally reproduced as they only require a dipole in the potential at the last scattering surface.  We draw our final discussions in section~\ref{sec:conclusions}. Finally, in 3 appendices we discuss in depth some of our derivations. In~\ref{appA} we derive the temperature anisotropies up to second order; in~\ref{appB1} we consider the second order perturbation relations that are relevant for our work; in~\ref{appB} we digress over the scalar-vector-tensor decomposition and the freedom we have at defining quantities at the origin.



\section{Aberration and second order CMB perturbations}\label{sec:2nd-order}

In what follows we will make use of the following notation: vectors and tensors will be written in boldface; their components will not be in boldface, and will carry Latin letters super-scripts which run from 0 (denoting the time component) to 3; second order perturbation quantities will always carry a subscript ``2''. Note however that by definition we will always treat vectors and tensors as in Euclidean space and so we can raise and lower indices in all such quantities just with a Kronecker delta; in fact, we take into account any effect due to the metric explicitly as extra terms in perturbation theory. For this reason we can use boldface notation without ambiguities, just to obtain a more compact notation. We assume a flat FLRW (Friedman-Lemaître-Robertson-Walker) background metric and that the gravitational theory is General Relativity. Also, we shall use units in which $c = 1$.

Consider the line element $\dd s^2=a^2(\eta) g_{\mu \nu} \dd x^\mu \dd x^\nu$, where $x^\mu=( \eta, \vr)$, $\eta$ is the conformal time and $a$ is the scale factor. The metric elements can be expanded up to second order in the following way:
\begin{align}
    g_{00}&\,=\, -\lp 1+2\phi+\phi_2\rp \,,%
    \label{eq:m1} \\
    g_{0i}&\,=\, z_i+\frac{1}{2} z_{2i} \,,%
    \label{eq:m2} \\
    g_{ij}&\,=\, \lp 1-2 \psi -\psi_2\rp \Od _{ij}+
    \chi_{ij}+\frac{1}{2} \chi_{2ij} \,,
    \label{eq:m3}
\end{align}
$\psi, \phi, \vz,\vc$ are first order perturbations to the background metric and  $\psi_2, \phi_2, \vz_2,\vc_2$ are second order terms, with $\vc$  traceless in order to make the separation of $g_{ij}$ unambiguous.

It is shown in Appendix \ref{appA} that for an observer with peculiar velocity $ \vo$ (see \eq{U}) observing photons coming from direction $ \vn$, the CMB temperature anisotropies
are given by
\begin{align} \label{dt/t}
\frac{\Od T(\vn)}{T} &=\vo \cdot \vn+T_1(\eta_*,\vr_*)+T_2(\eta_*,\vr_*)+\vo \cdot \vn\ T_1(\eta_*,\vr_*)+ \lp \vo \cdot \vn \rp^2\,,
\end{align}
where a subscript $*$ means evaluated at the Last Scattering Surface (LSS) or decoupling time, $\vr= (\eta_o- \eta) \vn$ describes the coordinates of a photon from the LSS to us, as inferred by the observer ignoring perturbations, and $\eta_o$ is the present conformal time.
Here $T_1$ contains the first order temperature fluctuations while $T_2$ contains second order terms:
\begin{align}
T_1 \,&=\, \phi-\vg\cdot \vn+\tau-\cI \,,\label{dt1} \\
T_2 \,&=\, \frac{\phi_2-\phi^2+ \tau_2}{2}+\f{\cI^2}{2}-\cI_2-\cI \lp \tau+\phi\rp+ \phi\ \tau-\vg \cdot \vn\ T_1 -\frac{\vg^2}{2}-\frac{1}{2}\vgg\cdot \vn+\delta T_p\,, \label{dt2}
\end{align}
where $\vg+\vgg/2$ is the three ``velocity'' of the emitter (photon-baryon fluid) which is related to the fluid four-velocity by (see Eqs.~\eqref{Ug}-\eqref{3-vel}). We introduce such a notation for the three-velocity following~\cite{Mirbabayi:2014hda,Creminelli:2011sq}, which also makes the equations more compact.
\begin{align} \label{3-vel-def}
\mathbf{v}&=\mathbf{u},  & &\mathbf{v}_2=\mathbf{u}_2 -2\mathbf{u}\psi+ \mathbf{u} \cdot\vc\,, \\
%
U_\gamma^\mu &= (U_\gamma^0,U_\gamma^i),   & &U_\gamma^i= \f{1}{a} \lc u_\gamma^i+\f12 u^i_{2\gamma}\rc\,.
\end{align}

The other quantities are defined below, but we note now that the very last term in $T_2$ contains the second order corrections due to the perturbed path characterized by $\vr+ \dr$:
\begin{align} \label{t1-expansion}
\delta T_p \,&=\, \dr \cdot \lc \nabla  \phi - (\nabla v_\gamma^i) n_i+ \nabla \tau\rc - \vg\cdot \lc \f{\vc \cdot \vn-\vn (\vn\cdot \vc\cdot \vn)}{2} +\dn\rc +\pp{\tau}{d^i} \delta d^i \nonumber \\
    &-\int_{\eta_o}^{\eta }d\bar\eta\ \lb \dr \cdot \lc \nabla \phi'+ \nabla \psi'+ \big(\nabla z_i'\big)n^i -\frac{1}{2}n^i \big(\nabla \chi_{ij}'\big) n^j\rc+ \vz'\cdot \dn - \vn\cdot \vc'\cdot \dn\rb\,,
\end{align}
where $\tau$ and $\tau_2/2$ are the first and second order intrinsic temperature anisotropies at the LSS, and $\p \tau/ \p d^i$ tells us how the emission of photons depends on the direction of emission $\mathbf{d}$ (in case of an anisotropic source), and the explicit form of $ \delta\mathbf{d}$ is given in \eq{d}. In this paper we assume that $ \tau$ is determined exclusively by the energy density perturbations of the photons (see~\eq{intinsic}) which depends only on position, and in that case $\p \tau/ \p d^i$ vanishes. Finally, a prime means derivative with respect to conformal time $\eta$ ($\p_0\equiv \ '\ \equiv \p_{\eta}$).

Some comments are in order. Eqs.~\eqref{dt1} and \eqref{dt2} were obtained in appendix \ref{appA} by rewriting previous results of~\cite{Mollerach:1997up} including scalar, vector and tensor contributions. They are fully general and valid in all gauges. This result is not new, except for two additions. First, we are introducing a new ingredient, which is necessary when comparing to observations: the direction of observation $\vn$ (seen by an observer with peculiar velocity $\vo$) instead of the unit vector $\ve$ (the direction of observation as seen by a comoving observer), following the notation of Ref.~\cite{Mollerach:1997up}.  In order to study second order perturbations we must relate both up to first order (see appendix \ref{appA} and \eq{ve-vo} in particular)
\begin{equation}\label{eq:e-vs-n}
    \ve \,=\, \vn + \lc - \vo + \vn ( \vo \cdot \vn) \rc \,.
\end{equation}
This will allow us to explicitly describe aberration. The other simplifying assumption we made is that $\psi=\phi=\vz=\vc=0$ at the observer's position, which we set as the origin. This is  always possible as we discuss in Appendix \ref{appB} and simplifies somehow our equations. Note also that in \cite{Mirbabayi:2014hda,Creminelli:2011sq} the authors have a similar formula for the case of primordial scalar perturbations, but which is valid only in the particular choice of the Poisson gauge.

We now define the remaining quantities appearing in eqs.~\eqref{dt1}-\eqref{t1-expansion}:
\begin{align}
    \cI&=\int_{\eta_o}^{\eta }d\bar\eta\ A' \,, \\
    A&= \phi+\psi+\vz\cdot \vn-\frac{1}{2}\vn\cdot \vc\cdot \vn\,,\\
    \cI_2&=\int_{\eta_o}^{\eta }d\bar\eta \lc \f{1}{2}A_2'+(2\psi'+ \vz'\cdot \vn- \vn \cdot \vc'\cdot \vn) A-(2 \phi+ \vz \cdot \vn) A'\rc , \label{I2}\\
    A_2&= \phi_2+\psi_2+\vz_2\cdot \vn-\frac{1}{2} \vn\cdot \vc_2\cdot \vn\,.
\end{align}
Here $\cI$ can be interpreted in the Poisson gauge as the integrated Sachs-Wolfe effect, and $ \cI_2$ as the Rees-Sciama effect.

When we observe a CMB photon, we see it as coming from the apparent emission point with coordinate $\vr_*$. However, the true photon emission point at the LSS is given by the coordinates  $\vr_*+ \dr_*$. Such a correction is given up to first order by
\begin{align}
    \dr \,&=\, \vn\ \delta{r} +\dr_\perp, \qquad\quad \Od r =- \int_{\eta_o}^{\eta } d\bar\eta\ A \,, \label{vre}\\
    \dr_\perp &= \lc -\vo+ \vn(\vo \cdot \vn) \rc ( \eta_o- \eta)
    -\int_{\eta_o}^{\eta } d\bar\eta\ \! \big[ \vz-\vn(\vz \cdot \vn) -\vc\cdot \vn+\vn (\vn \cdot\vc\cdot \vn)+(\eta -\bar\eta) \vA_\perp \big]\,, \label{dvrp}
\end{align}
where
\begin{align}
    \vA_\perp &=\nabla_\perp (\phi+\psi)+ (\nabla_\perp z_{i}) n^i-\frac{1}{2}n^i \lp \nabla_\perp \chi_{ij} \rp n^j, \\
    \nabla_\perp &= \nabla- \vn ( \vn \cdot \nabla )\,.
\end{align}

We want to stress that the true position at which a given photon was emitted is one and unique, in that sense $\vr_*+ \dr_*$ is observer independent. Since we have defined our zero order trajectory to be a straight line along the observed direction $\vn$, the apparent position of the CMB anisotropies represents an aberrated and lensed image of the sky, for that reason, the above formula provides aberration and lensing corrections. In fact, the velocity dependent terms in~\eq{dvrp} are \emph{de-aberration} terms, while, in the Poisson gauge, the integrated term could be called a \emph{de-lensing} term.  Finally, the second order anisotropies due to $ \delta r$ in~\eq{vre} can be interpreted in the Poisson gauge as time-delay.  We further discuss these quantities in section~\ref{sec:abe+len}.

Aberration and lensing give a total deflection angle $ \va$ which relates the direction of the true emitting point to the apparent one\footnote{See \eq{va1}, and also \cite{Lewis:2006fu}-pp.7-8 and \cite{Seljak:1995ve}-pp.4}
\begin{align} \label{va}
    \va \;=\; \f{1}{r}\dr_\perp(\eta, \vn)\,,
\end{align}
but there is also a local\footnote{By ``local'' we mean that it depends of each point on the trajectory, which is in contrast with global (or total) deflection angle. We borrow this terminology from \cite{Lewis:2006fu}-pp.7-8}  deflection angle given by\footnote{This notation agrees with Eq.(3.5) of~\cite{Mirbabayi:2014hda} except for the fact that we put aberration and lensing together.}
\begin{align} \label{dn}
    \dn = \lc -\vo +\vn ( \vo \cdot \vn) \rc + \vz- \vn( \vz \cdot \vn ) -\vc\cdot \vn +\vn (\vn \cdot\vc\cdot \vn )+\int_{\eta_o}^{\eta }d\bar\eta\ \vA_\perp\,.
\end{align}
Note that it is $ \va$ that is relevant in the Poisson gauge for the discussion of lensing effects on CMB, because one is interested in the total deflection (the angular excursion) of a photon as it travels from the LSS to our observation point, and not in the change in its direction (see~\cite{Seljak:1995ve}, pp.5).

In an analogous way to the discussion above, we see that $\vn+ \dn$ is observer independent, as can be seen when expressing $ \vn$ in terms of $ \ve$. It tells us that the photon 3-momentum is being deflected due to inhomogeneities from the LSS up to the origin, and is therefore observer independent, depending only upon the perturbations along the geodesic.

As stated before, the above formulas are fully general and valid in all gauges. But in what follows we will restrict ourselves to the case of pure scalar perturbations in the Poisson gauge.

\subsection{Poisson Gauge and Scalar Perturbations} \label{sec:poi+sca}

At first order each mode (scalar, vector and tensor) evolves independently and so we can study each one separately. To simplify our discussion, we consider only primordial scalar perturbations since vector modes are decaying and tensor modes are known to be subdominant compared to the scalar modes~\cite{Ade:2015tva}. However, at second order we need to consider every mode, since vector and tensor perturbations are sourced due to non-linearities by the first order scalar counterpart. We choose to work in the Poisson gauge which yields (see  \cite{Matarrese:1997ay,Bartolo:2005kv})
\begin{align}
    z_i=\chi_{ij}=0, \qquad \p^i z_{2i}=\p^i \chi_{2ij}=0\,.
\end{align}
It is also well known that in the Poisson gauge the first order potentials $\phi$, $\psi$ are equal (as long as we neglect anisotropic stress).


\section{Comparing aberration and lensing effects} \label{sec:abe+len}

In the first part of this work we discuss the second order temperature anisotropies due to lensing and aberration. We note that such terms appear explicitly into $ \dr_\perp$. However, it could also happen that aberration and lensing terms are hidden into intrinsic second order quantities, like $ \phi_2$. This is not the case for the Poisson gauge as we will stress later when discussing the Doppler-like effects (vide \eq{phi2}); however it can happen in other gauges as discussed in~\cite{Roldan:2015}.

Aberration effects appears explicitly when we compare two different frames, one of them in which the observer has, say $ \vo=0$, and the other in which $ \vo\ne 0$. That implies that the arrival direction of photons seems different in each frame. On the other hand, lensing is the effect due to gravitational perturbations along the photon's path, which also changes the apparent direction of emission. It is an integrated effect encoded into $ \dr_\perp $. Just for completeness we mention that $ \Od r$ describes the so called time-delay or Shapiro delay \cite{Hu:2001yq} which tells us that photons are not coming from a spherical shell of radius $r$ but from a distorted surface whose ``radius'' in direction $ \vn$ is $r+ \Od r$.

It is convenient to use spherical coordinates so that $\vr=(r,\vn)$ where $r$ is the radial coordinate centered at the observer position,
and we split the gradient into its radial and transverse parts
\begin{equation}
\nabla= \vn \der{}{r} + \f1r \wh, \qquad \wh=\hat{\theta}\ \p_\theta+\hat{\varphi} \f1{\sin \theta}\p_\varphi\,,
\end{equation}
so that $ \wh=r\nabla_\perp$ is the gradient on the unit $2$-sphere. With this definition we can see that
\begin{align} 
-\vo+ \vn(\vo \cdot \vn)=- \wh (\vo \cdot \vn),
\end{align}
and (see Eqs. \eqref{dvrp} and \eqref{va})
\begin{align}
\dr_\perp = -r\wh (\vo \cdot \vn)+r\wh \bar \phi, \Ra \va =\wh \lp -\vo \cdot \vn +\bar \phi\rp\,.\label{drpd}
\end{align}
Here we introduced the lensing potential
\begin{align}
\bar\phi(r, \vn)= 2\int_0^{r} d \bar r\ \f{\bar r-r}{\bar r r} \phi \,, \label{len-pot}
\end{align}
and used the fact that along the photon's geodesic we need to set $r=\eta_o-\eta$.

Note that lensing is similar to aberration; the difference is that in general $\bar\phi$ contains all terms of the multipole expansion (all the $\ell$'s and $m$'s), while $ \vo\cdot \vn$ has just the dipole $\ell=1$. This is related to the known properties of lensing and aberration: the former couples a large range of multipoles $\ell$ (see~\cite{Lewis:2006fu,Ade:2015zua}), while for the latter the most relevant coupling is between $\ell$ and its neighbors $\ell\pm 1$ (see~\cite{Challinor:2002zh,Notari:2011sb}).

For $ \dn$ we have
\begin{align}
\dn &= - \wh (\vo \cdot \vn)- 2\wh \int_0^{r}  \f{d\bar r}{\bar r} \phi\,,
\end{align}
which allows us to write the transverse part of \eq{t1-expansion} as\footnote{That is, the part due to $\dr _\perp$. We do not consider the radial part here, which corresponds to time-delays.}
\begin{align}\label{abe+len1}
    \delta T_{p\perp}&=\dr_\perp \cdot \lc \nabla  \phi - (\nabla v_\gamma^i) n_i+ \nabla \tau\rc - \vg\cdot \dn
    -2\int_{\eta_o}^{\eta }d\bar\eta\ \dr_\perp \cdot \nabla \phi' \notag \\
    &= \va \cdot \wh \lc \phi - \vg \cdot \vn+ \tau\rc
    -2\int_{\eta_o}^{\eta }d\bar\eta\ \va \cdot \wh \phi' +\vg \cdot \lp \va- \dn \rp \notag \\
     & \equiv T^{\rm deflec}+ \vg \cdot \nabla_\perp \delta r\,,
\end{align}
where we used the fact that $ \wh \lp \vg \cdot \vn \rp = (\wh v_\gamma^i) n_i + \vg \cdot \va$, and
\begin{align}
    \va- \dn \;=\; \f{2}{r} \, \wh \int_0^{r}  d\bar r\ \phi \;=\; \nabla _\perp \delta r\,.
\end{align}
Here $T^{\rm deflec}$, defined through \eq{abe+len1} contains both lensing and aberration. Note that in previous equations we defined $\va$ only at the observer, while $\dn$ is a vector defined at any point in the trajectory. However we remind the reader that we are treating all our vectors simply raising and lowering their (three-dimensional) indices with a Kronecker delta so we can safely sum such quantities together.

For the purpose of this work, we explicitly disentangle the modes that contribute to the dipole of the CMB from the rest. We proceed as follows: the gravitational potential can be expanded into spherical harmonics as $\phi( \eta, \vr )= \phi_{\ell m}( \eta,r) Y_{\ell m}( \vn)$, from which we can extract the dipolar part (\emph{i.e.} terms containing only $\ell=1$). We will need to consider the radial profile of $ \phi$ in order to compute its lensing potential and other integrated terms. However, for $ \vg$ and $ \tau$, we only need to consider their values at LSS. In the following we therefore find convenient to split each field as:
\begin{align} \label{split}
    \phi \to \phi_d+\phi,  \qquad \tau \to \tau_d+\tau,  \qquad \vg \to \vgd+ \vg \,.
\end{align}
In this way we can extract the dipolar contribution from $T_1$ and add it to $\vo \cdot \vn$ so that the total dipole of the CMB becomes\footnote{$T_2$ can also contain a second order contribution to the dipole but this is irrelevant in what follows.}
\begin{align} \label{td}
    \Theta_d \,=\, \vo \cdot \vn+\phi_d- \vgd\cdot \vn+\tau_d-2\int_{\eta_o}^{\eta }d\bar\eta\ \phi_d' \,.
\end{align}
In what follows we refer explicitly to all the remaining multipoles (\emph{i.e.} everything in $\ell>1$) coming from $T_1$ simply as ``$\Theta$'':
\begin{align}
    \Theta\,=\,\lp \phi-\vg \cdot \vn+\tau\rp-2\int_{\eta_o}^{\eta }d\bar\eta\ \phi'\,.
\end{align}
\eq{dt/t} can be thus written as
\begin{align} \label{T/T}
    \frac{\Od T(\vn)}{T} \,=\, \Theta_d+ \Theta+T_2+\vo \cdot \vn\ \Theta+ \textrm{quadrupole terms}\,.
\end{align}
In the following we do not consider  quadrupole terms explicitly. The reason is that we focus on aberration-like and Doppler-like couplings which affect all scales, so one can safely ignore the quadrupole terms with no effective loss of information.


\subsection{Aberration vs Dipolar Lensing}

We now explicitly separate the dipolar dependence from $T^{\rm deflec}$ and call it $T^{\rm deflec}_d$ (see \eq{abe+len1})
\begin{align}
    T^{\rm deflec}_d& \equiv \va_d \cdot \wh \lc \phi - \vg \cdot \vn+ \tau\rc
    -2\int_{\eta_o}^{\eta }d\bar\eta\ \va_d \cdot \wh \phi'\,, \label{gen-aberr} \\
    \va_d (\eta, \vn)&=\wh \lp -\vo \cdot \vn +\bar \phi_d\rp\,, \label{va_d}
\end{align}
which both describes aberration and dipolar lensing (lensing due to $ \phi_d$). \eq{gen-aberr} is the more general expression describing aberration+dipolar lensing; for clarity of discussion we now focus on the two extreme cases in which the dipole is purely of kinematical (boosted induced) or intrinsic (lensing induced) type.

\paragraph{Boosted induced dipole ---\!\!\!}
If the observed dipole is purely kinematical, we can set $\phi_d= \tau_d= \vgd=0$ as an extreme case, so $\Theta_d=\vo \cdot \vn$ and we see that the deflection is given by
\begin{align} \label{kin-aber}
    T^{\rm deflec}_d \,=\, -\wh ( \vo \cdot \vn ) \cdot \wh \lc \lp \phi-\vg \cdot \vn+\tau\rp -
    2\int_{\eta_o}^{\eta }d\bar\eta\ \phi' \rc \,=\, -\wh ( \Theta_d)\cdot \wh \Theta\,,
\end{align}
where we used the fact that $\vo \cdot \vn$ is constant along the line of sight, so that we could take it out of the integral.
This is a pure aberration term and it induces couplings between multipoles which grow linearly with $\ell$~\cite{Challinor:2002zh,Amendola:2010ty,Kosowsky:2010jm}.

\paragraph{Intrinsic Dipole ---\!\!\!}
We want now to check whether an intrinsic dipole can also produce the same couplings as aberration through the lensing effect.  In the more general case a given potential will induce a velocity to the observer which is proportional to $ \nabla \phi$ at the origin (see~\eq{velO}). But let us consider here the case $\vo=0$, so that the temperature anisotropies are totally determined by $\phi,\vg$ and $\tau$ (the general case is given in next section). We have
\begin{equation}
    \Theta_d=\lp \phi-\vg \cdot \vn+\tau\rp_d -2\int_{\eta_o}^{\eta }d\bar\eta\ \phi'_d\,,
\end{equation}
and the lensing due to $\phi_d$ is
\begin{align} \label{tad}
    T^{\rm deflec}_d &= \wh \lp \bar \phi_d \rp \cdot \wh \lp \phi-\vg \cdot \vn+\tau\rp -2\int_{\eta_o}^{\eta }d\bar\eta\  \wh \lp \bar \phi_d \rp \cdot \wh \phi'\,.
\end{align}
A primordial dipole will thus produce exactly the same couplings as a boost induced aberration if this satisfies the integral condition
\begin{align} \label{ic}
    T^{\rm deflec}_d  = -\wh ( \Theta_d)\cdot \wh \Theta\,.
\end{align}

Note that there exists no function $ \phi_d$ for which $ \bar \phi_d$ is a constant function of $r$ (or equivalently of $ \eta$, see~\eq{len-pot}), so we cannot take $ \wh \bar \phi_d$ out of the integral as we did in the boosted case. This expresses the fact that that lensing is cross correlated to the integrated Sachs-Wolfe effect (ISW). Also note that lensing is correlated with other effects like the Sunyaev-Zel'dovich effect (SZ) but for that we would need to consider the late reionization due to the hot gas. However, the SZ effect is not included in formula \eqref{dt/t}.  Generally, the lensing effect on the reionization and ISW signals is very small as they are only important on large scales (see \cite{Lewis:2006fu}-pp.6). Henceforth we will thus assume for simplicity full matter domination from recombination up to present time, which results in zero ISW effect. With such considerations we can then write \eq{gen-aberr} as
\begin{align} \label{app-aberr}
    T^{\rm deflec}_d \approx \va_d \cdot \wh \Theta\,,
\end{align}
and the above aberration-mimicking condition is
\begin{align} \label{ab-cond}
    \va_d =-\wh \Theta_d\,.
\end{align}
This would be the condition that insures that the aberration couplings measured by Planck \cite{Aghanim:2013suk} are consistent with the measurement of the CMB dipole.

\subsection{Matter Domination} \label{rad-phi}

The $0-i$ Einstein equation at first order in Poisson gauge is (see Eq.(149) of \cite{Bartolo:2004if})
\begin{align}
    \nabla\lp \psi'+\cH\phi\rp =-\f32 \cH^2 (1+w) \mathbf{v} \,,
\end{align}
where $w=P/\rho$, $ \mathbf{v}$ is the total-fluid velocity perturbation and $ \cH=a'/a$ is the conformal Hubble parameter. Here $P$ and $ \rho$ are the unperturbed pressure and density of the total fluid (matter+radiation). In matter domination $w=0$, $\phi = \psi$, $\phi'=0$ and $\cH=2/\eta$, so
\begin{align}\label{vel}
    \mathbf{v}\,=\, -\f23 \frac{\nabla}{\cH} \phi \,=\, -\f13\eta \nabla \phi  \,,
\end{align}
and we can compute the velocity induced by the potential $ \phi_d$ on the observer as
\begin{align}\label{velO}
    \vo^d=-\f13\eta_0 \nabla \phi_d(r=0) \,.
\end{align}
In general \eq{vel} only applies to the total matter fluid, however we can use it to compute $\vg$ on large scales. In principle one could imagine situations in which the photon fluid has a different large scale velocity compared to the rest, by invoking a non-standard initial condition or vector modes. However in the present paper we do not consider such possibilities, on the basis that anyway such modes would rapidly decay. Discarding such cases, then no causal processes such as free streaming or diffusion can separate the components, so all fluid velocities are equal (see~\cite{Hu:1994jd}-pp.8). Therefore on large scales the photon-fluid velocity is also given by \eq{vel}.\footnote{Note however that one could imagine exotic cosmologies with a very large scale photon fluid velocity, as an initial condition. Such a velocity would generically decay and in presence of Inflation it would then be forbidden. For these reasons we do not consider it further in the present paper.}

During matter domination the gravitational potential is constant in time, the integrated Sachs-Wolfe term $\cI$ vanishes, and the aberration-mimicking condition \eq{ab-cond} reduces to (see \eq{va_d} and \eq{td})
\begin{align} \label{aberr-cond-2}
\bar\phi_d( \vn,r_*)&= -\lc \phi_d- \vgd\cdot \vn+\tau_d \rc_* \,.
\end{align}
Note that in arriving at this equation we made no hypothesis on the observer's velocity, so it is valid even for nonzero $\vo$. So we can conclude that, if the lensing potential $\bar\phi_d$ satisfies the equation above, the Planck measurement~\cite{Aghanim:2013suk} in principle does not tell us which is the source of aberration, either of kinematic or intrinsic type. However there is no reason to expect that such a condition is satisfied and so one expects to see deviations between the velocity inferred from the CMB dipole and the one inferred through aberration, as long as $\phi_d$ is nonzero.

Now, in order to find radial profiles which satisfy~\eq{aberr-cond-2}, we need to discuss the intrinsic temperature anisotropies $ \tau$. They are given by (see Eqs.(80)-(81) of~\cite{Bartolo:2003bz})
\begin{equation}\label{intinsic}
    \tau=\f{1}{4}\Od_{\Og}, \qquad \tau_2= \f{1}{4}\Od_{2\Og}- 3\tau^2\,,
\end{equation}
where $\Od_{\Og}$ and $\Od_{2\Og}$ are the first and second order density-contrast perturbations of the photons. If $ \varrho$ is a energy density, with mean value $ \rho$, then density-contrast perturbations are defined by
\begin{align}
    \varrho=\rho \lp 1+\Od+\f12 \Od_2+\cdots\rp, \qquad
    \Od=\f{\Od \varrho}{\rho}, \qquad \Od_2=\f{\Od \varrho_2}{\rho}\,.
\end{align}
In order to be more general let us consider also the possibility of having isocurvature (or entropy) perturbations. We introduce the entropy perturbation between the total matter fluid (cold dark matter + baryonic matter) and radiation:
\begin{align}
    S=3\lp \f{\Od_m}{3}-\f{\Od_\Og}{4}\rp\,.
\end{align}
Using the fact that on large scales $\Od_m=-2 \phi$ (see \eq{delta-phi}) we can rewrite the integral condition as
\begin{align} \label{aberr-cond-3}
\bar\phi_d( \vn,r_*)&= -\lc \f{\phi_d-S_d}{3}+ \f{\eta}{3} \der{\phi_d}{r}\rc_* \,,
%
\end{align}
where we used the relation
\begin{align}\label{velg}
    \vgd \cdot \vn \;=\; -\f{\eta_*}{3} \vn \cdot\nabla \phi_d\;=\; -\f{\eta_*}{3} \der{}{r}\phi_d ( \vn,r_*)\,.
\end{align}
Note that we are considering the possibility of an isocurvature perturbation only for the very large scales which correspond to a dipolar potential, so usual constraints from CMB do not apply and moreover on such large  scales we can treat baryons and CDM as a single fluid.
Now, that for a given radial profile $\phi_d(r)$, we can always satisfy \eq{aberr-cond-3} by just choosing the appropriate initial condition for $S_d$. However, again there is no reason in principle to expect that such a condition is exactly satisfied. Thus we regard such a choice for $S$ as a fine tuning of the initial conditions. We come back to this issue in the end of this Section.

The velocity term $\vgd$ is suppressed by a factor of $\sim \eta_*/r_* \approx \eta_*/ \eta_o$, that is, by a factor of about 100 compared to $\phi$, so we neglect it in the following. We now consider two extreme cases: pure adiabatic perturbations $S=0$, and pure initial isocurvature perturbations, for which $\,\phi_*=-S/5\,$ (see \eq{phi-AS}). In addition, we choose the $z$-axis so that $\phi_d(\vr)=\phi_d(r) \cos\theta$. The integral condition can thus be written as (see \eq{len-pot})
\begin{align}\label{phid}
    \phi_d(r_*)&\,=\, N\int_0^{r_*} dr\ \phi_d (r) \lp \f1 r-\f{1}{r_*}\rp .  \qquad
     \begin{cases}
    N=1,   &  \textrm{Isocurvature}\\
    N=6,   &  \textrm{Adiabatic}
    \end{cases}
\end{align}

Note that we study here the two extreme cases of pure adiabatic or isocurvature, but more generally $N$ can take other values by taking linear combinations of them.
Given a radial profile $\phi_d(r)$ we can check if it produces aberration-like couplings with the appropriate magnitude or not by just computing the above integral. It is also clear that for a fixed $r_*$ there exist an infinite number of functions $ \phi_d$ which do satisfy \eq{phid}. However in general, if they do satisfy the condition for a given $r_*$ they will do not for another value of the LSS radius. In this sense, though we can mimic aberration with an intrinsic dipole, it would be in principle a fortuitous situation. Note also that from the infinitely many solutions to \eq{phid} there exist one which is  valid for any value of $r_*$. In fact consider $R=r_*$, and take two derivatives with respect to $R$ in both sides to obtain
\begin{align}
    R^2\derr{}{R}\phi_d(R)+2R\der{}{R}\phi_d(R)=N \phi_d(R)\,,
\end{align}
or
\begin{align}
    \phi_d(R)=
     \begin{cases}
    c_1 R^{-3}+c_2 R^2,   &  \textrm{Adiabatic}\\
    c_1 R^{m_1}+c_2 R^{m_2},   &  \textrm{Isocurvature}
    \end{cases}
\end{align}
with $c_1,c_2$ constants and $m_{1,2}=(-1\pm \sqrt{5})/2$. The condition $\phi_d(R=0)=0$ sets $c_1 =0$, then we finally get
\begin{align}
    \phi_d(R)\propto
     \begin{cases}
     R^2,   &  \textrm{Adiabatic}\\
     R^{(-1\pm \sqrt{5})/2}\,.   &  \textrm{Isocurvature}
    \end{cases}
\end{align}
The isocurvature solution is not an analytical function of $R$ as its derivative diverges at the origin. Since we expect the gravitational potential to be analytical, we can conclude that isocurvature modes can satisfy the aberration-mimicking condition only for fixed values of $r_*$. Adiabatic perturbations on the other hand have a solution $ \phi_d(r) \propto r^2$, which is valid for all $r_*$. Clearly, this solution must have this parabolic form only in the region $0\le r\le R$ for some $R \gtrsim r_*$; the functional form outside the horizon can be of any type, since we are in unobservable regions.

We summarize this section by saying that in order to reproduce Planck \cite{Aghanim:2013suk} measurements on aberration couplings, a dipolar potential has to be negligible with respect to our peculiar velocity (as is usually assumed) or it has to satisfy to the integral condition \eq{aberr-cond-3}. In the purely kinematic case, this integral constraint is automatically satisfied as $\phi_d=0$. However, there are some radial profiles which satisfies the constrain independently of the amplitude of the perturbation. Since the dipole of CMB only fixes the amplitude of the field at decoupling, a primordial dipole could in principle still produce aberration-like couplings just as a kinematical dipole. Nonetheless, we do not expect the dipolar potential to satisfy this integral condition, as such, a possible discrepancy between the inferred peculiar velocity from a kinematic dipole and aberration could be accounted for by a non-negligible contribution from a dipolar potential.

There is still one radial profile for adiabatic perturbations which mimics aberration without any fine tuning in the amplitude or the LSS radius. However it seems that in that case the observer is located at a special position in the Universe, one in which the radial profile seems locally as $ \phi_d(r) \propto r^2$ and the angular profile a $\cos(\theta)$. Unless there is some mechanism that comes to justify this radial profile, this also seems as a fine tuning (reminiscent of the one found in inhomogeneous void models for dark energy~\cite{Alnes:2005rw,Blomqvist:2009ps,Quartin:2009xr}).

Even in the case in which the radial profile could mimic a kinematic dipole, there might nevertheless be a way to completely disentangle lensing from aberration by looking at the cross correlation between a dipolar lensing and the ISW (in a more realistic scenario, including dark energy) could help to solve this degeneracy as a kinematic aberration has no correlation with the ISW. The same could perhaps be done by looking at cross correlation between a dipolar lensing and SZ. We leave a more quantitative investigation for a future work.


\section{Doppler-induced multipole couplings} \label{sec:doppler}

The observer's peculiar velocity contributes to the CMB dipole through $\vo \cdot \vn$, but it also induces Doppler couplings given by $(\vo \cdot \vn) \ \Theta$, (see \eq{T/T}). If the dipole is purely kinematical then $\Theta_d= \vo \cdot \vn$, and the total Doppler effect is
\begin{align} \label{T-dop}
    T^{\rm dop}( \vn)= \Theta_d\ \Theta\,.
\end{align}
Planck \cite{Aghanim:2013suk} measurements are consistent with the Doppler formula given above, so this can be seen as a consistency check for a kinematical dipole.

However, we now show that an intrinsic dipole also leads to Doppler-like terms. For this purpose, we turn our attention to the terms inside $T_2$ besides $\delta T_p$ (see \eq{dt2}). We dub these collection of terms ``$\Theta_2$''. We still assume full matter domination and take the large scale limit so that we can neglect terms involving $ \vg$ and $\vgg$ ($\vgg$ is subdominant as we can see by looking at the $0-i$ second order Einstein equation, the same way as we did with $ \vg$). So, in this case $ \Theta_2 \equiv T_2- \delta T_p$ becomes
\begin{align} \label{t2-t2d}
\Theta_2 + \Theta_{2d} & =  \frac{\phi_2-\phi^2+ \tau_2}{2}-\cI_2+ \phi\ \tau+ \lc \frac{\phi_2-\phi^2+ \tau_2}{2}+ \phi\ \tau-\cI_2 \rc_d + \textrm{quad. terms}\,,
\end{align}
where we have used the splitting given in \eq{split} and ``quad. terms'' means contributions to the quadrupole. Here, the subscript $d$, means the part of that expression which is linear in the dipole quantities. So for example,
$(\phi^2)_d \equiv 2\phi\ \phi_d$.

To proceed we need to know what are the expressions for $\phi_2$ and $\psi_2$. It is shown in appendix \ref{app:iso+adia} that\footnote{See also Eq.(3.14) of \cite{Mollerach:1997up}, Eqs.(2.18)--(2.26) of \cite{Bartolo:2005kv} and Eq.(B.4) of \cite{Bartolo:2006fj}.}
\begin{align}
    \phi_{2}&=2\phi^2-6\cK +\lnl+\f1{14} \f{ \p_i \p_j}{ \nabla^2}\lp \f{10}{3}\p_i \phi \p_j \phi- \delta_{ij}\p^k \phi \p_k \phi  \rp \eta^2\,,\label{phi2}\\
    \psi_{2}&=-2 \phi^2+4\cK+\lnl+\f1{14} \f{ \p_i \p_j}{ \nabla^2}\lp \f{10}{3}\p_i \phi \p_j \phi- \delta_{ij}\p^k \phi \p_k \phi  \rp \eta^2\,,\label{psi2}\\
    \cK &= \f{ \p_i \p_j }{ \nabla^4} \lc \p_i \phi \p_j \phi- \f{\delta_{ij}}{3} \p^k \phi \p_k \phi \rc\,,
\end{align}
where summation over repeated indexes is assumed. Here, $\lnl$ is associated with the presence of primordial non-Gaussianity and it is given by
\begin{align} \label{ini-adia}
\lnl &=\f35 \lp -\zeta_2+2\zeta^2 \rp \,, \qquad  \quad \textrm{adiabatic},\\
\lnl &=\f15 \lp -S_2+\f23 S^2\rp \,, \qquad \textrm{isocurvature}\,,
\end{align}
where $ \zeta, \zeta_2$ ($S, S_2$) are first and second order curvature (isocurvature) perturbations. It is important to stress that in the adiabatic case $\zeta_2$ is usually parametrized as $\,\zeta_2 = 2a_{\rm nl} \zeta^2\,$ so that $ \lnl=-5(a_{\rm nl}-1) \phi^2/3$, as was done e.g. in~\cite{Bartolo:2005kv,Mirbabayi:2014hda}. Here, however, we take a more general approach and make no assumption on the initial conditions for $\zeta_2$.

In Eqs. \eqref{phi2}-\eqref{psi2}, $\eta \to 0$ at the very beginning of matter domination which occurs at some time after recombination. Note that since we are considering the simplified case in which the Universe is matter dominated from the LSS until today, then we can set $ \eta \to 0$ in the second order quantities when evaluating at LSS.

In the remaining of this Section we will compute for simplicity only the perturbations on large scales, but we have reason to believe that all the results below can also be extended to small scales as well. We will pursue a careful derivation of this extension in a future work. This is important observationally, as most of the Doppler (and aberration) couplings signal comes from these scales~\cite{Notari:2011sb}.


\subsection{Adiabatic perturbations}

For adiabatic perturbations and on large scales we have (see \eq{tau-adia})
\begin{align} \label{tau_d-adia}
\tau=-\f23 \phi, \qquad \Theta=\f13\phi, \Ra \phi_d\ \tau +\phi\ \tau_d=-\f23 (\phi)^2_d\,,
\end{align}
while (see \eq{tau2-adia})
\begin{align}
\tau_2= \lp \frac{16\phi^2-6\phi_{2} }{9} \rp_d  \,.
\end{align}
Here, $\phi$ and $\phi_{2}$ have to be evaluated at LSS. After replacing those expression into $\Theta_{2d}$ (see \eqref{t2-t2d}) we get\footnote{This agrees with Eq. (1.2) of \cite{Bartolo:2005kv}.}
\begin{align}\label{T2d-3}
\Theta_{2d}&=\f12\lc \phi_{2m}-\phi^2+\lp \frac{16\phi^2-6\phi_{2m}}{9}\rp \rc_d -\cI_{2d}
 -\f23 (\phi)^2_d \notag\\
&=\lp \f{\phi^2}{18}\rp_d-\cK_d+\frac{1}{6}\lnl-\cI_{2d}\,.
\end{align}
Remembering that $ \phi+ \tau= \phi/3$, then $\lp \phi^2\rp_d=2 \phi\ \phi_d=18 \lp \phi+ \tau \rp_d \Theta$, and so
\begin{align}\label{t2d-adia}
\Theta_{2d}= \Theta\ \lp \phi+ \tau \rp_d -\cK_d+\frac{1}{6}\lnl-\cI_{2d}\, .
\end{align}
%


\subsection{Isocurvature perturbations}

For  isocurvature perturbations, we have on large scales (see \eq{tau-iso})
\begin{align} \label{tau_d-iso}
    \tau=\phi, \qquad \Theta=2\phi, \Ra \phi_d\ \tau +\phi\ \tau_d=(\phi)^2_d\,,
\end{align}
while (see \eq{tau2-iso})
\begin{align}
    \tau_2= \lp \psi_2+3\phi^2 \rp_d  \,.
\end{align}
Replacing into $\Theta_{2d}$ yields
\begin{align}
\Theta_{2d}&=\f12\lc \phi_{2m}-\phi^2+ \lp \psi_{2m}+3\phi^2 \rp \rc_d -\cI_{2d}+ (\phi)^2_d\notag\\
&= 2\lp \phi^2\rp_d-\cK_d+\lnl-\cI_{2d}\,.
\end{align}
Remembering that $ \phi=\tau$, leads to $2\lp \phi^2\rp_d=4 \phi\ \phi_d=\lp \phi+ \tau \rp_d \Theta$, then
\begin{align}\label{t2d-iso}
\Theta_{2d}= \Theta\ \lp \phi+ \tau \rp_d -\cK_d+\lnl-\cI_{2d}\,.
\end{align}
We need now to add the kinematic contribution, which leads to
\begin{align} \label{doppler}
\vo \cdot \vn\ \Theta+\Theta_{2d}= \Theta_d\ \Theta + \f{\lnl}{N}-\cK_d-\cI_{2d},  \qquad
\begin{cases}
N=1,   &  \textrm{Isocurvature}\\
N=6,   &  \textrm{Adiabatic}
\end{cases}
\end{align}
where we used that the total dipole is $\Theta_d=\vo \cdot \vn +\lp \phi+ \tau \rp_d$. Note that this $N$ already appeared in the aberration discussion. As a consistency check note that such a product $ \Theta_d\ \Theta$ is a generalization of Eq.~(4.11) of ~\cite{Mirbabayi:2014hda}, which was derived in the special case of a linear gradient mode (which is a special case of our dipolar potential) and in absence of primordial non-Gaussianity and which agrees with our result.

We now parametrize $\lnl$ as
\begin{align}
    \lnl=-\f53(a_{\rm nl}-1) \phi^2\,,
\end{align}
not only for the adiabatic case (as was done in \cite{Bartolo:2005kv,Bartolo:2006fj}) but also for the isocurvature case. Taking the $d$-dependent part
\begin{align}
    (\lnl)_d=- \f{10}{3}(a_{\rm nl}-1) \phi\ \phi_d\,,
\end{align}
and remembering that $\Theta=\phi/3$ for adiabatic and $\Theta=2\phi$ for isocurvature perturbations, we get the Doppler-like couplings given by the initial conditions
\begin{align} \label{doppler-ini}
    T_{\rm dopp-i.c.}=\f{\lnl}{N}= - \f{5}{3}(a_{\rm nl}-1) \phi_d \Theta.
\end{align}
Now, from \cite{Bartolo:2005kv}-pp.(5) we know: in the standard scenario $ \anl \approx 1$, while in the curvaton case $ \anl = 3/(4r) - r/2$, where $r$ is the relative curvaton contribution to the total energy density at curvaton decay. In the minimal picture for the inhomogeneous reheating scenario,
$ \anl=1/4$.

Eqs. \eqref{t2d-adia}, \eqref{t2d-iso} and \eqref{doppler-ini} are our main results about Doppler-like couplings. If $\lnl~\approx~0$, as predicted by single field inflation, then remarkably we see that such equations lead exactly to the same couplings as the one produced in the usual kinematic Doppler case. This striking result shows that in the simplest inflationary scenario there is no way to disentangle a large scale potential from a boost using only Doppler-like couplings. In other words: suppose that a future experiment will measure a deviation in the values of $ \vo$ inferred from the dipole and from the Doppler-like couplings. This would imply that there must be a large dipolar potential perturbation and moreover coming from a non-standard inflationary mechanism which generates non-Gaussianity.

We note that a dipolar potential also produces other effects which are due to $\cK_d+\cI_{2d}$. This is similar to what happens when studying aberration, where a $\phi_d$ produces other effects which a peculiar velocity does not, e.g, time-delay. As can be seen $\cK$ contains non-local terms (due to the inverse Laplacians); however, as discussed in \cite{Mirbabayi:2014hda} (Section 5),  those non-localities must cancel somehow as observations cannot depend on perturbations well outside our observable Universe. This is explicitly shown in \cite{Mirbabayi:2014hda} (Appendix C), where they noted that the non-locality of $ \cK$ is cancelled by boundary terms coming from the integral $ \cI_2$. However other terms survive, contributing to the CMB anisotropies. This can be see explicitly from
\begin{align}
    \cK+\cI_2= \f{1}{3} \int_{\eta_o}^{\eta } d\bar\eta\ \eta \lc 2 ( \vn \cdot \nabla  \phi)^2- \nabla \phi \cdot \nabla \phi \rc\,,
\end{align}
which follows from Eq.(C.1) of \cite{Mirbabayi:2014hda}. We see therefore, that the additional terms coming from $\cK_d+\cI_{2d}$ do not contribute to aberration-like or Doppler-like couplings. Finally, we stress that $\cK_d+\cI_{2d}$ is an integrated term, so it can in principle be correlated with  with dipolar lensing, and also with SZ and ISW effects.

\section{Conclusions}\label{sec:conclusions}

The Planck satellite has detected couplings between multipoles in the CMB at all scales which are consistent with Doppler and aberration effects due to our peculiar velocity~\cite{Aghanim:2013suk}. In particular the Doppler couplings are $\ell$-independent, while the aberration ones grow linearly with $\ell$~\cite{Challinor:2002zh,Amendola:2010ty}. The measured values are consistent in amplitude and direction with the well-known measured CMB dipole. However in this paper we have tried to check whether a large scale dipolar gravitational potential could produce or not the same observational signatures. We have shown that such a potential, in addition to a dipole in the CMB, produces indeed couplings due to lensing, which are similar to aberration, and couplings due to second order potentials, which are similar to Doppler. However, crucially, we find important differences compared to a boost effect. We illustrate in Table~\ref{tab:summary} our main results, which we discuss in more detail below.

\begin{table}
\begin{center}
    \begin{tabular}{lccc}
    \hline
    \hline
    & $10^{-3}$ dipole & $10^{-8}$ Doppler-like& $10^{-8}$ aberration-like \\  & & couplings & couplings \\
    \hline
    Peculiar velocity & yes  & yes & yes \\[4 pt]
    Adiab.~dipolar potential & yes & yes${}^\star$  & only with fine-tuning  \\[4 pt]
    Isocur.~dipolar potential & yes  & yes${}^\star$  & only w/ \emph{even more} fine-tun. \\[4 pt]
    Non-Gauss. dipolar pot. & yes  & different & only with fine-tuning\\[2 pt]
    \hline
    \hline
    \end{tabular}
\end{center}
\vspace{-.25cm}
\caption{
    Summary of the main conclusions of this paper. The ${}^\star$  symbol is a reminder that we have only been able to prove the corresponding result on large scales. See Section~\ref{sec:conclusions} for details.
\label{tab:summary}}
\end{table}%

First, the amplitude of the lensing couplings discussed above, relative to the dipole, depends on the radial profile of the dipolar potential and in most cases only for fine-tuned choices one could exactly reproduce a boost effect. For the adiabatic case a parabolic radial profile could exactly mimic aberration without any fine tuning in the LSS radius. However, the observer needs to be located at a special position in the Universe, to wit near the center of such parabola. Unless there is some mechanism that comes to justify this, this can be seen as a fine-tuning. For the isocurvature case, the problem is even worse as one would need also to fine-tune the distance to the LSS.

So generically we expect to see a deviation in the measured value of $\vo$ through such couplings due to the possible presence of a dipolar potential. One expects of course such a potential to be present at least with a $10^{-5}$ amplitude, but perhaps higher in more exotic models which violate the global isotropy of the Universe. Detecting a larger deviation would signal the presence of an unusually large dipolar potential.

Second, we have found that a primordial dipolar potential can also reproduce Doppler-like couplings (just as a kinematical dipole) both for an adiabatic and isocurvature dipole, as shown in \eq{t2d-iso} and \eq{t2d-adia}. Such an effect depends on the initial conditions, presumably set by Inflation. Strikingly in this case a single-field slow-roll scenario would induce couplings which are {\it exactly} degenerate with a boost. So, detecting {\it any} inconsistency between the value of $\vo$ from the CMB dipole and the one from the Doppler couplings would signal the presence of physics beyond single-field inflation.  As stated before, for the Doppler-like couplings we derived the equations only on the large scale regime. We have reason to believe that this result might extend also to small scales. We will develop a more careful proof in a future work. If this is true, the only approximation here would be to assume we are always in the complete matter domination regime.

We stress that a primordial dipole produces in addition other effects which are potentially detectable: cross correlation with the ISW and SZ effects, time-delay and integrated effects coming from $\cK_d+\cI_{2d}$, and which we will discuss in detail in a future paper.

In conclusion, better measurements of the dipole, the Doppler and aberration couplings could in principle help us distinguish a boost from a dipolar gravitational potential; the latter however can be detected only if it has an unusually large amplitude compared to the other CMB multipoles.


\emph{\textbf{Acknowledgments.}}
We thank Filippo Vernizzi and Maur\'icio Calv\~ao  for useful discussions. MQ and OR are grateful to Brazilian research agencies CNPq and FAPERJ for support. AN is supported by the grants EC FPA2010-20807-C02-02, AGAUR 2009-SGR-168.

\appendix
\import{\.}{appendix-JCAP-vFinal.tex}

\bibliographystyle{JHEP2015}
\bibliography{aberration}

\end{document}

%% file: appendix-JCAP-vFinal.tex

\section{Derivation of temperature anisotropies up to second order}\label{appA}

In this appendix we obtain Eqs.~\eqref{dt/t}-\eqref{t1-expansion} by interpreting and rewriting some previous results of \cite{Mollerach:1997up}. For clarity we repeat here the metric given at the beginning of this work,
\begin{align}
g_{00}&=-\lp 1+2\phi+\phi_2\rp\,, \label{g00}\\
g_{0i}&=z_i+\frac{1}{2} z_{2i}\,,\\
g_{ij}&=\lp 1-2 \psi -\psi_2\rp \Od _{ij}+
\chi_{ij}+\frac{1}{2} \chi_{2ij}\,, \label{gij}
\end{align}
but this time we use coordinates $x^\mu=( \eta, \vx)$ for an easy comparison with \cite{Mollerach:1997up}. We will also make use of the unit vector $\ve$, defined as
the direction of observation for a comoving observer. We then define the functions
$T_1=T_1(\eta,\vx)$ and $\cT_2=\cT_2(\eta,\vx)$ as
\begin{equation}\label{dt1app}
T_1=\phi-\ug \cdot \ve+\tau-\cI,
\end{equation}
and
\begin{align}
    \cT_2 &=\frac{\phi_2-\phi^2}{2}-I_2-(\ug\cdot \ve) \ \phi+
    \lp \cI+\ug\cdot \ve \rp \lp - \uo\cdot \ve - \phi- \tau+\ug\cdot \ve+\cI \rp \nonumber\\
    &+ \delta x^0 A'+(\dx+\delta x^0 \ve) \cdot \lc \nabla \phi- (\nabla u_\gamma^i) e_i+ \nabla \tau\rc +\frac{ \uo^2- \ug^2}{2}+\phi\ \tau+\pp{\tau}{d^i} \delta d^i  \nonumber\\
    &+\uo\cdot \ve \lp \tau+\phi\rp-\ug \cdot \lp \vz + \vi\rp +\frac{1}{2}( \mathbf{u}_{2o}-\ugg)\cdot \ve+\f12\tau_2,\label{dt2app}
\end{align}
where each quantity is defined below in Eqs.~\eqref{I1}-\eqref{d}. Mollerach and Matarrese \cite{Mollerach:1997up} (see also \cite{Pyne:1995bs}) have shown that the CMB anisotropies at first and second order are given by\footnote{Note that we have changed a bit the notation of \cite{Mollerach:1997up}, see Eqs.~(2.4)--(2.7) of that work for comparison. E.g. we interchange $ \phi$ with $ \psi$.}
\begin{align} \label{t12}
    \frac{\Od T_1( \ve)}{T} &=\uo\cdot \ve+T_1(\eta_*,\vx_*), \qquad  \frac{\Od \cT_2( \ve)}{T}=\cT_2(\eta_*,\vx_*),\qquad \vx_* \,=\,  \lp \eta_o- \eta_* \rp  \ve \,,
\end{align}
where $*$ means a quantity evaluated at decoupling (sometimes referred in the literature with an $\cE$, for ``emission'') and $\eta_o$ the present conformal time.~\eq{t12} describes the CMB anisotropies as seen by an observer with four velocity
\begin{align} \label{U}
    U^\mu_o = (U^0_o,\mathbf{U}_o),  \qquad \mathbf{U}_o = \mathbf{u}_o+\f12 \mathbf{u_{2}}_o\,,
\end{align}
which measures the frequency of arrival photons with ``normalized'' four-momentum
\begin{align} \label{K}
k^\mu = \der{x^\mu}{\eta}= (k^0, \vk).
\end{align}
At the observer, we have
\begin{align} \label{Kobs}
 k^\mu_o \,=\, (1,-\ve).
\end{align}

Some comments are in order: in the previous equations we set $\psi=\phi= \vz=\vc=0$ at the observer's position (the origin), this is always possible as we discuss in Appendix \ref{appB}. Second, we have included two additional terms $( \mathbf{u}_{2o}-\ugg)\cdot \ve/2$ and $\tau_2/2$ which were not considered in the final expressions of  \cite{Mollerach:1997up} but were included in more recent papers~\cite{Bartolo:2004ty}.

We now define the quantities appearing in Eqs.~\eqref{dt1app} and \eqref{dt2app}.
\begin{align}
\cI&=\int_{\eta_o}^{\eta }d\bar\eta\ A',
& &\vi=\int_{\eta_o}^{\eta }d\bar\eta\ \vA,\label{I1} \\
A&= \phi+\psi+\vz\cdot \ve-\frac{1}{2}\ve\cdot \vc\cdot \ve,
& &
\vA= \nabla (\phi+\psi)+ (\nabla z_{i}) e^i-\frac{1}{2}e^i (\nabla \chi_{ij}) e^j, \label{vA}\\
\delta k^0  &=-2\phi-\vz\cdot \ve+\cI, & &\delta \vk=-2 \psi \ \ve-\vz +\vc \cdot \ve-\vi, \label{k0}\\
\delta x^0&=\int_{\eta_o}^{\eta } d\bar\eta \lc-2\phi-\vz\cdot \ve+(\eta -\bar\eta)A'\rc,
& &\dx=-\int_{\eta_o}^{\eta } d\bar\eta \lc 2 \psi\  \ve+\vz-\vc \cdot \ve+(\eta -\bar\eta) \vA\rc,\label{x}
\end{align}
where a prime means derivative with respect to conformal time $\eta$ ($\p_0=\ '\ =\p_{\eta}$). Note that all quantities defined in the LHS of Eqs.~\eqref{I1}-\eqref{x} are scalars while those of the RHS are vectors. Then we have
\begin{align}
I_2&=\int_{\eta_o}^{\eta }d\bar\eta \lp\frac{1}{2} A_2 +\cA_2\rp,  \qquad  \qquad
 A_2= \phi_2+\psi_2+\vz_2\cdot \ve-\frac{1}{2} \ve\cdot \vc_2\cdot \ve,\\
\cA_2&=-( \vz'-\vc'\cdot \ve)\cdot (\delta \vk+ \ve\ \delta k^0)+2\ \delta k^0 A'+2\psi' A+\delta x^0 A^{''}+\dx \cdot \vA'.\label{A3}
\end{align}

Finally, $\ug$ and $\ugg/2$ are perturbations to the velocity of the emitter (photon-baryon fluid) at first and second order, $\tau$ and $\tau_2/2$ are the first and second order intrinsic temperature anisotropies at the LSS and
$\p \tau/ \p d^i$ tells us how the emission of photons depends on the direction of emission (due to possible anisotropic emissions), with
\begin{align} \label{d}
\delta \mathbf{d}=\ve-\frac{\ve-\delta \vk}{| \ve-\delta \vk |}\,.
\end{align}

In general, the four-velocity is defined by
\begin{align} \label{Ug}
U^\mu = (U^0,\mathbf{U}),  \qquad \mathbf{U}= \f{1}{a} \lc \mathbf{u}+\f12 \mathbf{u}_2\rc \,.
\end{align}
Here we use instead a related quantity,  with which the equations become  more compact:
\begin{align}
\mathbf{V}= \f{1}{a} \lc \mathbf{v}+\f12 \mathbf{v}_2\rc,  \qquad
\mathbf{V}=\mathbf{U} (1- \psi)+ \mathbf{U} \cdot \f{ \vc}{2} \,.
\end{align}
Then we have the relations
\begin{align} \label{3-vel}
\mathbf{v}=\mathbf{u},  \qquad \mathbf{v}_2=\mathbf{u}_2 -2\mathbf{u}\psi+ \mathbf{u} \cdot\vc \,.
\end{align}
Note that this redefined three-velocity satisfies
\begin{align}
\delta_{ij}V^iV^j&= \lp U^i (1- \psi)+\f12 U_j \chi^{ji} \rp \lp U_i (1- \psi)+ \f12 U^k\chi_{ki} \rp \notag \\
 &= (1-2 \psi)U^i U_i+ U^i \chi_{ij} U^j+ \mathcal{O}(4)= g_{ij} U^i U^j+\mathcal{O}(4)\,.
\end{align}


\subsection{Direction of observation and deflection angles}

We draw the attention on the fact that $ \ve$ is not in general the true direction of observation. To see this, note that the direction of observation $n^\mu$ of a photon is defined through\footnote{In the Eqs.~\eqref{U}, \eqref{K} and \eqref{N} we are setting $a=1$ as we are at the observer's position.} (see~\cite{King:1972td})
\begin{align} \label{N}
k^\mu = \omega (U^\mu-n^\mu),  \qquad \omega=- \gmn k^\mu U^\nu.
\end{align}
Plugging Eqs.~\eqref{g00}-\eqref{gij}, and \eqref{U}-\eqref{K} into \eq{N}, we arrive at
\begin{align} \label{ve-vo}
    \ve&= \vn \lp 1+ \uo \cdot \vn+ \vn \cdot \f{ \vc_o}{2} \cdot \vn - \psi_o \rp - \uo \notag \\
     & = \vn \lp 1+ \uo \cdot \vn\rp - \uo\,,
\end{align}
where the subscript $o$ means evaluated at the observer's position.\footnote{In \cite{Mirbabayi:2014hda} they set $ \vc_o=0$ from the beginning but they missed the term $\psi_o$ into direction vector $ \vn$. See Eq.(A.6) of \cite{Mirbabayi:2014hda}.} The first line of the previous equation is a general result, in the second line we have used that in our case $\psi_o=\vc_o=0$. As the notation suggest, $\delta k^0$ and $\delta \vk$ are perturbations to the photon's four momentum, they are first order and as we see from the previous formulae they generate second order perturbations so that we do not need to consider higher order perturbations of $ \vk$. The same applies to $\delta x^0$ and $\delta \vx$ which are first order perturbations to the photon's path produced by inhomogeneities along the journey from the LSS to us. Note that we do not include second orde terms in~\eqref{ve-vo}, since they would only affect third order observables.

It is helpful to split the photon's path perturbation into its radial and transverse part, as $\,\dx+ \delta x^0\ve= \ve\ \Od x +\dx_\perp\,$
with
\begin{align} \label{dx}
    \Od x &=(\dx \cdot \ve + \delta x^0)=
    -2\int_{\eta_o}^{\eta } d\bar\eta \lc \phi+\psi+ \vz\cdot \ve - \ve \cdot \f{ \vc}{2}\cdot \ve -
    \f{\eta -\bar\eta}{2} \der{A}{ \eta}\rc \notag \\
     & = (\eta- \bar \eta) A\Big|_{ \eta_o}^{\eta}-2\int_{\eta_o}^{\eta } d\bar\eta \lp A- \f{A}{2} \rp =- \int_{\eta_o}^{\eta } d\bar\eta\ A\,,
\end{align}
where in the first line we used the fact that
\begin{align} \label{der}
    \der{}{ \tau} = \p_0 - \ve \cdot \nabla,
\end{align}
and in the second line we integrated by parts and used $A=0$ at origin. Now,
\begin{align} \label{dx-perp}
\dx_\perp &=
-\int_{\eta_o}^{\eta } d\bar\eta \lb \vz- \ve( \vz \cdot \ve ) -\vc\cdot \ve +\ve (\ve \cdot\vc\cdot \ve )+
(\eta -\bar\eta)  \vA_\perp\rb \,,
\end{align}
where the definition of $\vA_\perp$ is the same as that of $ \vA$ (see \eq{va}) but using the transverse gradient $\nabla_\perp$ instead
\begin{align} \label{n_perp}
    \nabla_\perp \equiv \nabla- \ve ( \ve \cdot \nabla )\,.
\end{align}
In the same manner we split $\dk+ \delta k^0\ve= \ve\ \Od k +\dk_\perp$ with
\begin{align}
    \Od k  & =- A\,, \\
    \dk_\perp  & =-\lc \vz- \ve( \vz \cdot \ve ) -\vc\cdot \ve +\ve (\ve \cdot\vc\cdot \ve )+ \int_{\eta_o}^{\eta }d\bar\eta\ \vA_\perp \rc\,.\label{de}
\end{align}
%


Note that at zero order $e^i=-k^i$ represents both the backward direction of the photon's momentum, and also the direction of observation for an observer with zero peculiar velocity, $\uo=0$. However, at first order, those concepts are not degenerate. Then what is the meaning we should give to $ \de$? Hereafter, we define
\begin{align} \label{de-def}
\de \equiv -\dk_\perp\,,
\end{align}
it therefore will represent the local deflection angle due to lensing and should not be confused with the total defection angle $\va$ given by\footnote{See \cite{Lewis:2006fu}, pp. (7-8) and \cite{Seljak:1995ve}, pp.4. Note that our notation for $\de$ is different from theirs.}
\begin{align} \label{va1}
\va= \f{1}{r}\dx_\perp\,,
\end{align}
which relates the direction of the true emitting point to the apparent one. \\

As noted before, $\vx= ( \eta_o- \eta) \ve$ describes the unperturbed path followed by photons from LSS to us, while the perturbed path is characterized by $ \vx+ \dX$:
\begin{equation} \label{dX}
\dX \equiv \dx+ \delta x^0\ve =\ve\ \Od x +\dx_\perp\,,
\end{equation}
now, take the function $T_1$ evaluated at the ``true'' emission point $\vx+ \dX$ and compare with its value at the apparent emission point $\vx$,
\begin{align} \label{t1-exp}
\delta \bar T &\equiv T_1( \eta, \vx+\dX)- T_1( \eta, \vx)= \delta\phi - \lc (\delta u_\gamma^i) e_i+\ug\cdot \de \rc +\delta\tau\notag \\
-&\int_{\eta_o}^{\eta }d\bar\eta\
\lc \delta\phi'+\delta\psi'+ \lp (\delta z_i')e^i +\vz'\cdot \de\rp -\lp  \frac{1}{2}e^i \delta\chi_{ij}' e^j+\ve\cdot \vc'\cdot \de\rp  \rc\,,
\end{align}
where in the last line we use the fact that $ \vc$ is symmetric. Then, we find
\begin{align} \label{deltat1}
\delta \bar T&= \dX \cdot \lc \nabla  \phi - (\nabla u_\gamma^i) e_i+ \nabla \tau\rc - \ug\cdot \de+\pp{\tau}{d^i} \delta d^i\notag \\
&-\int_{\eta_o}^{\eta }d\bar\eta\
\lb \dX \cdot \lc \nabla \phi'+ \nabla \psi'+ (\nabla z_i')e^i -\frac{1}{2}e^i (\nabla \chi_{ij}') e^j\rc+ \vz'\cdot \de - \ve\cdot \vc'\cdot \de\rb\,,
\end{align}
where we used
\begin{align} 
\delta\phi=\dX \cdot \nabla \phi,  \qquad \delta\psi=\dX \cdot \nabla \psi,  \qquad \delta \vc=\dX \cdot \nabla \vc,
\end{align}
and
\begin{align} \label{delta-tau}
\delta\tau=\dX \cdot \nabla \tau+\pp{\tau}{d^i} \delta d^i\,.
\end{align}

Here, we took into account that in general the emission of photons at LSS depends not only on position but may also depend on the direction. The direction of emission $ \mathbf{d}$ do change by a quantity $ \delta \mathbf{d}$ when we go from the unperturbed path to the perturbed one (see \eq{d}).


\subsection{Isolating the contributions from the perturbed path }

We now proceed to rewriting \eq{dt2app}, it allow us to get the simple form given a the beginning of this work. First note that, by using $ \p_0=d/d\tau + \ve \cdot \nabla $, we can rewrite the term $\delta x^0 A^{''}$ as
\begin{align}
\delta x^0 A^{''}=\delta x^0 \lp \der{A'}{ \eta}+\ve \cdot\vA' \rp,
\Ra \delta x^0 A^{''}+\dx \cdot \vA'=\delta x^0 \der{A'}{ \eta}+ \dX \cdot \vA',
\end{align}
now using
\begin{align}
\int_{\eta_o}^{\eta }d\bar\eta\ \delta x^0 \der{A'}{\eta}=
\delta x^0 A^{'} \Big|_{ \eta_o}^{\eta} -\int_{\eta_o}^{\eta }d\bar\eta\ \delta k^0 A^{'},
\end{align}
and the fact that the fields vanish at origin, $\delta x^0 A^{'} \Big|_{ \eta_o}=0$, yields
\begin{align}
\cT_2 &=\frac{\phi_2-\phi^2+ \tau_2}{2}- \tilde I_2-(\ug\cdot \ve) \ \phi+\lc -\cI+(\uo- \ug)\cdot \ve \rc T_1 \nonumber\\
&+ \dX \cdot \lc \nabla  \phi - (\nabla u_\gamma^i) e_i+ \nabla \tau\rc-\frac{\ug^2}{2}+\phi\ \tau+\pp{\tau}{d^i} \delta d^i
-\ug \cdot \lp \vz + \vi\rp -\frac{1}{2}\ugg\cdot \ve,
\end{align}
where we neglect $ \uo^{2}$ and $ \mathbf{u}_{2o} \cdot \ve$ from $T_2$ as they only affect the monopole and dipole, respectively. $\mathbf{u}_{2o}$ can also be absorbed into a redefinition of $ \uo$. Here,
\begin{align}
\tilde I_2 & =\int_{\eta_o}^{\eta }d\bar\eta
\lc \f{1}{2}A_2'-( \vz'-\vc'\cdot \ve)\cdot ( \ve\delta k- \de)+\delta k^0 A'+2\psi' A+ \dX \cdot \vA'\rc\,,
\end{align}
where we used the relation $\de=- \dk_\perp$. Now, by using $ \ve \cdot \nabla =\p_0-d/d\tau$, we have
\begin{align}
\vz+\vi&=\vz+\int_{\eta_o}^{\eta }d\bar\eta \lc  \ve \lp \p_0-\der{}{ \tau} \rp A+ \vA_\perp\rc  \notag \\
& = \vz+\ve\lp \cI- A\Big|_{ \eta_o}^{\eta} \rp +\int_{\eta_o}^{\eta }d\bar\eta \vA_\perp \notag \\
& = \ve \cdot\lp \cI- \phi- \psi+\ve\cdot \frac{ \vc}{2}\cdot \ve\rp + \vz- \ve( \vz \cdot \ve )+\int_{\eta_o}^{\eta }d\bar\eta \vA_\perp \notag \\
& = \ve \cdot\lp \cI- \phi- \psi-\ve\cdot \frac{ \vc}{2}\cdot \ve\rp + \vc \cdot \ve+ \de\,.
\end{align}
So, by using \eq{deltat1} we get
\begin{align}
\cT_2 &=\frac{\phi_2-\phi^2+ \tau_2}{2}- \tilde\cI_2+\lc -\cI+(\uo- \ug)\cdot \ve \rc T_1+ \phi\ \tau\nonumber\\
&+\delta \bar T-\frac{\ug^2}{2}-\frac{1}{2}\ugg\cdot \ve+\ug \cdot \ve \lp -\cI+ \psi+\ve\cdot \frac{\vc}{2}\cdot \ve\rp - \ug \cdot \vc \cdot \ve,
\end{align}
with
\begin{align} \label{calI2}
\tilde\cI_2 & =\int_{\eta_o}^{\eta }d\bar\eta \lc \f{1}{2}A_2'+(z'\cdot \ve- \ve \cdot \vc'\cdot \ve) A+\delta k^0 A'+2\psi' A\rc \notag \\
 & = \f{\cI^2}{2}+\int_{\eta_o}^{\eta }d\bar\eta \lc \f{1}{2}A_2'+(2\psi'+z'\cdot \ve- \ve \cdot \vc'\cdot \ve) A-(2 \phi+ \vz \cdot \ve) A'\rc \notag \\
  & \equiv \f{\cI^2}{2}+ \cI_2 \,,
\end{align}
where we used \eq{k0} and the fact that $\int_{\eta_o}^{\eta }d\bar\eta\ \cI A'=\cI^2/2$. \\

We summarize this section as follows: the CMB temperature anisotropies up to second order are given by
\begin{align} \label{dT}
\frac{\Od T(\ve)}{T} &= \vo \cdot \ve+T_1(\eta_*,\vx_*)+T_2(\eta_*,\vx_*)+\vo \cdot \ve\ T_1(\eta_*,\vx_*), 
\end{align}
with $T_1$ given in \eq{dt1app}
\begin{align}\label{T2}
T_2 &=\frac{\phi_2-\phi^2+ \tau_2}{2}+\f{\cI^2}{2}-\cI_2-\cI \lp \tau+\phi\rp+ \phi\ \tau-\vg\cdot \ve\ T_1\nonumber\\
&-\frac{\vg^2}{2}-\frac{1}{2}\vgg\cdot \ve+\f12\vg \cdot \lc \ve (\ve\cdot \vc\cdot \ve) -\vc \cdot \ve \rc + \delta \bar T,
\end{align}
and the integral $\cI_2$ defined in \eq{calI2}. Here, and in the rest of this work, we used the three velocities $ \mathbf{v}, \mathbf{v}_2$ instead of $\mathbf{u}, \mathbf{u}_2$, see \eq{3-vel}.

Finally, we remember that for $\delta \bar T$ we must use \eq{deltat1}, with $ \dX$ given in Eqs.~\eqref{dX}, \eqref{dx-perp}  and \eqref{dx}; and $ \de$ given in \eq{de-def}. This is so due to our interpretation of $ \de$ as being the local deflection angle instead of total defection angle (see discussion before \eq{va1}). The expressions above together with \eq{ve-vo} allow us to easily obtain Eqs.~\eqref{dt/t}-\eqref{t1-expansion}.


\section{Second order gravitational potentials and intrinsic anisotropies} \label{appB1}

This appendix provides the formulas and concepts needed to study the Doppler couplings in Section~\ref{sec:doppler}. Though in that Section we explicitly separate the $d$-dependent part (dipolar components) in the potentials, in this appendix for simplicity $\phi, \phi_2, \cdots$ represent the full perturbations containing also the dipolar terms. In what follows, all equation are given in Poisson gauge.

The derivation that we follow in this appendix is similar to what was done in~\cite{Bartolo:2003bz}; we generalize it here by including also isocurvature perturbations and finally by specializing to the case of an explicit separation of the dipolar part in the potentials. \\

By using the traceless part of the $i-j$ Einstein's equations, one obtains (see \cite{Bartolo:2005kv}-Eqs.~(C.3-C.7), or \cite{Bartolo:2004if}-Eq.(154))
\begin{align}
\psi_2- \phi_2 &=-4 \phi^2+3 \f{ \p_i \p_j}{ \nabla^4} \lc F_{ij}- \f{\delta_{ij}}{3} F^k_{\ k} \rc\,, \\
F_{ij} &= 2\phi_i \phi_j+3 (1+w) \cH^2 v_i v_j,
\end{align}
where $ \mathbf{v}$ is the total fluid velocity, and summation in repeated indexes is assumed. Here for simplicity, we used $ \phi_i \equiv \p_i \phi$. The $0-i$ Einstein equation at first order (see \cite{Bartolo:2004if}-Eq.(149))
\begin{align} \label{euler-eq}
\nabla\lp \psi'+\cH\phi\rp =-\f32 \cH^2 (1+w) \mathbf{v} \,,
\end{align}
yields $ \phi_i=-3 \cH v_i/2$ in matter domination (hereafter MD), and $ \phi_i=-2\cH v_i$ in radiation domination (hereafter RD). From this we get $F_{ij} = 10\phi_i \phi_j/3$ for MD, and $F_{ij} = 3\phi_i \phi_j$ for RD, so
%
%
%
\begin{align}
&\psi_2- \phi_2 =-4 \phi^2+10 \cK, \qquad \textrm{MD} \label{psi2-phi2-MD}\\
&\psi_2- \phi_2 =-4 \phi^2+9 \cK, \qquad \ \ \textrm{RD} \label{psi2-phi2-RD}\\
&\cK \equiv \f{ \p_i \p_j}{ \nabla^4} \lc \phi_i \phi_j- \f{\delta_{ij}}{3} \phi^k \phi_k \rc.
\end{align}

Now, by taking the trace of the $i-j$ Einstein's equations, one obtains during MD (see \cite{Bartolo:2006fj}-Eq.(B.3))
\begin{align}
\psi_2''+ 3 \cH \psi_2'=\f{\p_i \p_j}{ \nabla^2}\lp \f{10}{3}\phi_i \phi_j- \delta_{ij}\phi^k \phi_k  \rp\,,
\end{align}
from which follows, with $ \cH=2/ \eta$
\begin{align*}
\phi_{2}&= \phi_{2m}+\f1{14} \f{ \p_i \p_j}{ \nabla^2}\lp \f{10}{3}\phi_i \phi_j- \delta_{ij}\phi^k \phi_k  \rp \eta^2\,,
\end{align*}
where $\phi_{2m}$ stands for the value of $\phi_2$ at the very beginning of MD, taken conventionally at $ \eta \to 0$ in \cite{Bartolo:2006fj}. In this appendix, we rewrite the equation above as
\begin{align}\label{phi2-eta}
\phi_{2}&= \phi_{2m}+\f1{14} \f{ \p_i \p_j}{ \nabla^2}\lp \f{10}{3}\phi_i \phi_j- \delta_{ij}\phi^k \phi_k  \rp (\eta^2-\eta_m^2)\,,
\end{align}
where $\eta_m$ is the value of $ \eta$ at the very beginning of MD.

\subsection{Initial conditions: Isocurvature and Adiabatic perturbations} \label{app:iso+adia}

The purpose of this section is to provide the tools for computing $ \Theta_{2d}$ as defined in \eq{t2-t2d}. In that sense, we are interested in terms which are linear in $ \phi_d$ or its gradient. In what follows we will find terms involving products of $ \phi_d$ with $ \phi$, or gradients of them. Then, it is useful to note that given the product of two arbitrary fields $f$ and $g$, we have in Fourier space
\begin{align}
\lp f\ g \rp( \vk)= \int d^3 \vk_1 d^3 \vk_2\ \Od( \vk_1+ \vk_2- \vk)\ f( \vk_1)\ g( \vk_2)\,.
\end{align}
%
Given an arbitrary mode $k$ of the product $f\ g $, it will in general involve all modes  of both $f$ and $g$. However, since in our case we have products of the form $f_d\ g$ where $f_d$ is already a large scale mode (small $k_1$), then $k$ and $k_2$ must be associated to the same scale. That is, $g$ determines the scale of the product $f_d\ g$.\footnote{Here $g$ stands either for $ \phi$ or $ \nabla \phi$, while $f_d$ stands for $\phi_d$ or $ \nabla \phi_d$.}

The $0-0$ Einstein equation at first order is (see \cite{Christopherson:2011ra}-Eq.(3.45), or \cite{Bartolo:2004if}-Eq.(148))
\begin{align}
\cH (\psi'+\cH\phi)-\f{1}{3}\nabla^2\psi=-\f12 \cH^2 \Od\,,
\end{align}
where $\Od$ is the total-fluid density-contrast perturbation. So, for large scales where we neglect gradient terms we have
\begin{align} \label{delta-phi}
    \Od=-2 \lc \f{\psi'}{ \cH}+\phi\rc  \,.
\end{align}
Below, we will need to apply this equation in two special cases: matter domination ($\Od=\Od_{m}$) and at the very beginning of radiation domination ($\Od=\Od_{ \gamma}$). In both cases we have
\begin{align}
\Od_m &=-2 \phi_m,  \quad\;\; \textrm{MD} \label{dm}\\
\Od_\gamma &=-2 \phi_p,  \qquad \textrm{RD,  \;  initial conditions,} \label{dg}
\end{align}
where we used $ \psi'=0$ at MD and the fact that $ \psi'$ is analytical so that $ \eta\, \psi' \to 0$ as $ \eta \to 0$.  In the radiation era we have $\cH=1/\eta$. In addition we use $\phi_p$ to denote the primordial value of the gravitational potential (at the very beginning of RD) while $\phi_m$ is the value in MD.

The $0-0$ second order Einstein equation is (see \cite{Christopherson:2011ra}-Eq.(3.129) or \cite{Bartolo:2004if}-Eq.(153))
\begin{align*}
3\cH (\psi_2'+\cH\phi_2)-\nabla^2\psi_2-3(\psi')^2-3\nabla\psi \cdot \nabla\psi-8\psi\nabla^2 \psi -12\cH^2\psi^2
=-3 \cH^2 \lc \f{\Od_2}{2} +(1+w) v^2\rc\,,
\end{align*}
where $\Od_2$ is the total-fluid density-contrast perturbation. Taking the $d$-dependent part (terms linear in $ \phi_d$ or $ \nabla \phi_d$), and the large scales limit we get
\begin{align} \label{2ener-const}
\cH (\psi_2'+\cH\phi_2)_d-(\psi')^2_d-4\cH^2 (\phi^2)_d=-\f12 \cH^2 \Od_{2d}\,.
\end{align}

We stress that in general, even when taking the large scale limit, we should not neglect terms like $\nabla\psi \cdot \nabla\psi$, because they will involve all scales of each individual field. However, as mentioned above when one of the scales is kept to be large (here $ \phi_d$) then necessarily the other quantity must also be a large scale. Combining this with $\mathbf{v} \propto \nabla \phi$ (see \eq{euler-eq}), we can safely neglect those terms.

We will henceforth, for simplicity, omit the subscript $d$ just to keep the notation cleaner. \eq{2ener-const} gives rise to
\begin{align}
\Od_{2m} &=-2\phi_{2m}+8\phi_m^2,  \quad\; \textrm{MD} \label{dm2}\\
\Od_{2 \gamma}&=-2\phi_{2p}+8\phi_p^2,  \qquad \textrm{RD,  \; initial conditions,} \label{dg2}
\end{align}
where we used again the analyticity of the potentials and \eq{phi2-eta}, from which follows that $\psi_{2}' \Big|_{ \eta_m}=0$.


\subsubsection{First order}

The continuity equation leads to
\begin{align}
\f{\Od'}{3(1+w)}+\f{\nabla \cdot \mathbf{v}}{3}-\psi' =0\,,
\end{align}
which is valid for any fluid, photons, baryons, CDM as they do not interchange
energy (baryons and photons do interchange momentum through Thomson scattering but not
energy, \cite{Lyth:2009zz}-pp.(130)). At large scales we can neglect $ \mathbf{v}$ so that, after
integration we have
\begin{align} \label{cont-eq}
\f{\Od_\Og}{4}-\psi =C_1, \qquad \f{\Od_m}{3}-\psi =C_2\,,
\end{align}
where we used the fact that for matter $w=0$, and for radiation $w=1/3.$ Here $C_1( \vx)$ and $C_2( \vx)$ are constant fields fixed by the initial conditions, the way we choose them defines if we are in the adiabatic mode, in the isocurvature mode or in a mixture of them. The adiabatic condition is that for which $C_1=C_2$, while the matter isocurvature, stands for the initial condition $\Od_\Og=\psi=0$ and therefore
$C_1=0$.

\paragraph{Adiabatic Case: $C_1=C_2$}

This condition immediately yields, using \eq{dm}
\begin{align} \label{tau-adia}
\tau\equiv  \f{\Od_\Og}{4}\Bigg|_{LSS}=\f{\Od_m}{3}\Bigg|_{LSS}=-\f23 \phi_m\,.
\end{align}

\paragraph{Isocurvature Case:}

The isocurvature case is that with initial contitions $\Od_\Og \Big|_p=\phi_p=C_1=0$. So \eq{cont-eq} yields
 $\Od_\Og=4\psi$ at any time (on large scales), then
\begin{align} \label{tau-iso}
\tau\equiv  \f{\Od_\Og}{4}\Bigg|_{LSS}= \phi_m\,.
\end{align}

\subsubsection{Second order}

The continuity equation at second order on large scales is\footnote{See also Eqs.~(4.9) and (9.4) of \cite{Bartolo:2006fj} for continuity equations at all scales of both radiation and CDM. See also Eq.(237) of~\cite{Bartolo:2004if}.}
\begin{align}
\f{\Od'_2-(\Od^2)'}{3(1+w)}-\lp \psi_2+2\psi^2 \rp' =0\,,
\end{align}
which is valid for any fluid only at large scales, because outside the horizon each
fluid evolves independently and no causal effect can produce energy transfer.
Integrating we get
\begin{align} \label{2cont-eq}
\f{\Od_{2\Og}-\Od_\Og^2}{4}-\lp \psi_2+2\psi^2 \rp =c_1, \qquad \f{\Od_{2m}-\Od_m^2}{3}-\lp \psi_2+2\psi^2
\rp=c_2\,.
\end{align}
Again, the initial conditions define in which mode we are.

\paragraph{Adiabatic Case: $c_1=c_2$}

This condition  leads to:
\begin{align}
\f{\Od_{2\Og}}{4} = \f{\Od_{2m}-\Od_m^2}{3}+\f{\Od_\Og^2}{4}=\f{1}{3}\Od_{2m}+\f{1}{9}(\Od_{m})^2\,,
\end{align}
where we used the adiabatic condition at first order. Using Eqs.~\eqref{dm2} and \eqref{dm} we arrive at
\begin{align} \label{tau2-adia}
\tau_2= \f{1}{4}\Od_{2\Og}\Bigg|_{LSS}- 3\tau^2 =
\lp \frac{-2\phi_{2m}+8\phi^2_m}{3}+\f49 \phi^2_m\rp -3\lp \f49 \phi^2_m\rp=\frac{16\phi^2_m-6\phi_{2m}}{9}\,,
\end{align}
where in the last line we use the assumption that recombination occurs at the beginning of matter domination so we can set $\phi_{2m}=\phi_{2*}$.

\paragraph{Isocurvature Case:}

The isocurvature case is that with initial contitions $\Od_{2\Og} \Big|_p=\psi_{2p}=c_1=0$.
So \eq{2cont-eq} yields at any time (on large scales)
\begin{align} 
\f{\Od_{2\Og}-\Od_\Og^2}{4}=\psi_2+2\psi^2, \qquad \Ra \f{\Od_{2\Og}}{4}=\psi_2+6\psi^2 \,,
\end{align}
where we used $\Od_\Og=4\psi$ from the first order isocurvature condition. This leads to
\begin{align}\label{tau2-iso}
\tau_2= \f{1}{4}\Od_{2\Og}\Bigg|_{LSS}- 3\tau^2 =\psi_{2m}+3\psi_m^2\,.
\end{align}

\subsection{Initial conditions: entropy and curvature perturbation}

The initial conditions are usually given in terms of the gauge invariant quantities like the curvature perturbation, which at first and second order are defined by\footnote{\eq{zeta2} follows from \cite{Malik:2008im}, pp.(41-44) and Eq.(6.3) of~\cite{Bartolo:2006fj}. }
\begin{align}
\zeta_i &= \f{ \delta_{i}}{3(1+w_i)}- \psi, \\
\zeta_{2i} &= \f{ \delta_{2i}}{3(1+w_i)}- \psi_2- \lc \f{1+3w_i}{(1+w_i)^2} \f{\delta_i^2}{9}+\f{4 \psi\ \delta_i}{3(1+w_i)}+\f{2 \delta_i}{9(1+w_i)} \f{ \nabla \cdot \mathbf{v}}{ \cH}\rc \,, \label{zeta2}
\end{align}
here $i$ stands for the type of fluid, either radiation or matter (baryons + cold dark matter). In terms of them it is usually defined the entropy (or isocurvature) perturbations
\begin{align}
S &= 3 (\zeta_m -\zeta_{\gamma}), \\
S_2 &= 3(\zeta_{2m}-\zeta_{2\gamma}).
\end{align}

On large scales we neglect the velocity dependent term that appears into the definition of $\zeta_{2i}$, and from the previous subsection we see that $\zeta_\gamma=C_1$, $\zeta_m=C_2$, while by using Eqs.~\eqref{2cont-eq} and \eqref{2cont-eq} we get
\begin{align}
    \zeta_{2\gamma}=c_1+2C_1^2,  \qquad \zeta_{2m}=c_2+2 C_2^2\,.
\end{align}
We see therefore that the curvature perturbations defined above are constant on large scales. We now relate $\phi_{m}=\psi_{m}$, $\phi_{2m}$ and $\psi_{2m}$ to the initial conditions.

\paragraph{First order}

Let's relate $\phi$ to its primordial value $\phi_p$. Using Eqs.~\eqref{dm}-\eqref{dg} into \eq{cont-eq} yields
\begin{align}
\zeta &\equiv \zeta_{\gamma}\Big|_{RD}= -\f32 \phi_p,  \qquad  \zeta_m = -\f53 \phi_m \label{Z}\,,
\end{align}
from which
\begin{align} \label{phi-AS}
\phi_m=
\begin{cases}
-\f35 \zeta= \f{9}{10} \phi_p,  \qquad \textrm{Adiabatic,} \\
-\f15 S,  \qquad  \qquad  \quad  \textrm{Isocurvature.}
\end{cases}
\end{align}

\paragraph{Second order}

From \eq{2cont-eq} we get
\begin{align}
3c_2&=\lc \Od_{2m}-\Od_m^2-3\lp \psi_{2m}+2\psi_m^2 \rp  \rc,  \qquad \Ra \notag \\
3 \lp \zeta_{2m}-\zeta_{m}^2 \rp &=\lc -2 \phi_{2m}-3\psi_{2m}-2\psi_m^2 \rc\,,
\end{align}
where we used Eqs.~\eqref{dm} and \eqref{dm2}. Now, by using \eq{psi2-phi2-MD} into the previous equation we get
\begin{align} \label{non-gauss1}
5\phi_{2m}-10\phi_m^2 +30 \cK &=
\begin{cases}
-3 \zeta_2+6\zeta^2\,, \qquad \textrm{adiabatic}, \\
-S_2+\f23 S^2\,, \qquad \textrm{isocurvature}.
\end{cases}
\end{align}
%
Here we set $\zeta_2\equiv \zeta_{2m}=\zeta_{2 \gamma}$ for adiabatic perturbations\footnote{\eq{non-gauss1} agrees with \cite{Bartolo:2005kv}-Eq.(2.17) for the adiabatic case.}. This previous result together with \eq{psi2-phi2-MD} immediately yields
\begin{align}
\phi_{2m}&=2\phi^2-6\cK + \lnl \,, \label{phi2app}\\
\psi_{2m}&=-2 \phi^2+4\cK+ \lnl \,, \label{psi2app}
\end{align}
with $\lnl$ as defined in \eq{ini-adia}. Those expressions together with \eq{phi2-eta} leads to Eqs.~\eqref{phi2}-\eqref{psi2}.

\section{Gauges and Scalar, vector and tensor perturbations.}\label{appB}

It is common to classify the perturbations as being of scalar, vector and tensor type according to their transformation behavior under rotation on spatial three hypersurfaces (see \cite{Matarrese:1997ay, Malik:2008im,Christopherson:2011ra}).
For simplicity, consider just the first order metric perturbation, then we have that $\phi$ and $\psi$ are scalars while we can write
\begin{align}
z_i&= \p_i z_s +z_i^{v},\\
\chi_{ij}&= \lp\p_i\p_j-
\f13\,\delta_{ij}\nabla^2\rp
\chi_s+\lp \p_i \chi_j^v+\p_j \chi_i^v\rp+\chi_{ij}^T\;,
\end{align}
where $z_s$ and $\chi_s$ are scalars, $z_i^v$ and $\chi_i^v$ are divergence free vectors; $\p^i z_i^v=\p^i \chi_i^v=0$, and $\chi_{ij}^T$ is a traceless and transverse tensor, that is $\lp\chi^T\rp^i_i=\p^i\chi_{ij}^T=0$.

Under a gauge transformation defined by the parameters $\xi^\mu=(\alpha,\xi^i)$ the metric changes at first order as follows~\cite{Malik:2008im,Matarrese:1997ay}
\begin{align}
\tilde{\phi}&=\phi+\alpha'+\cH\,\alpha \label{GT-phi}\,,\\
\tilde{z}_i&=z_i-\alpha_{,i}+\xi_i' \,,\\
\tilde{\psi}&=\psi-\f13 \xi^k_{,k}-\cH \,\alpha \,,\\
\tilde{\chi}_{ij}&=\chi_{ij}+\xi_{i,j}+\xi_{j,i}-\f23 \Od_{ij}\xi^k_{,k}\label{GT-chi}\,.
\end{align}
%

\subsection{Setting Minkowski at the origin}

Note that, the second order perturbations $\vz_2, \vc_2$ enters only in the integrated terms
through their derivatives, so that any shift like: $\vz_2\to \vz_2+constant$ in those functions
does not alter the observed anisotropies. Also, $\phi_2, \psi_2$, evaluated at the observer
position only affects the monopole, therefore we can safely set $\phi_2=\psi_2= \vz_2=\vc_2=0$
at origin. Let's now look at the first order quantities.

It is well known that for any spacetime we can always choose the coordinates in such a way that the metric looks locally Minkowski. However, in cosmological perturbation theory we work with a fixed background, in our case, the FLRW space-time, which is built as the solution for a
perfectly homogeneous fluid of density $\rho=\rho(\eta)$. So, setting the perturbations equal to zero at a given position, seems to indicate that we are supposing that the density at that point is equal to the mean density
of the Universe. But what if that is not the case? The question is that as far as we are in perturbation theory any point has a density which is not too different from the mean density. Since we are interested only on the observable Universe we can always superpose an isotropic and homogenous family of super-horizon modes\footnote{Those modes will belong to our background as seen in our observable Universe.} which cannot have any physical effect inside the horizon, but which will shift the metric perturbations by a small constant value. So by using this shifting we can set those perturbations to vanish at a desired point\footnote{Note however that this argument do not allow us to set the first derivative to zero.}. In the following we show it explicitly by using gauge transformation.

Start with a generic metric perturbation without fixing the gauge. Now, choose an arbitrary event $p$. We will show that we can always set the metric perturbation to vanish at that point.
We will do it in two steps.

\subsection*{(i) Set $ \vz=0$ and $ \vc=0$ at $p$} First, apply the gauge transformation given by
\begin{align}
\alpha=0, \qquad \xi^i= \f{ \omega_{ij}}{2} x^j + \Omega^i \eta\,,
\end{align}
for an arbitrary constant matrix $\omega_{ij}$. Then by using Eqs.~\eqref{GT-phi}-\eqref{GT-chi} we see that $z_i$ and $\chi_{ij}$ transform as
\begin{align}
\tilde{z}^i&=z^i+\Omega^i\,,\\
\tilde{\chi}_{ij}&=\chi_{ij}+\omega_{ij}-\f13 \Od_{ij}\ \omega^k_{\ k}\,.
\end{align}

By properly choosing $\omega_{ij}$ and $ \Omega^i$ we can get rid of $\vc$ and $ \vz$ at the desired point $p$.

\subsection*{(ii) Set $ \phi=0$ and $ \psi=0$ at $p$}

We now apply a transformation
\begin{align}
\alpha=\alpha( \eta), \qquad \xi^i= \omega x^i\,,
\end{align}
for some constant $\omega$. We see that $\vz$ and $ \vc$ become invariant, so that they still vanish at $p$, while
\begin{align}
\tilde{\phi}&=\phi+\alpha'+\cH\,\alpha \,,\\
\tilde{\psi}&=\psi- \omega-\cH \,\alpha \,.
\end{align}
By properly choosing $\alpha( \eta)$, we can set $\phi=0$ at $p$. After that, we can fix $ \omega$ to set $\psi=0$ at $p$.